\documentclass[12pt]{article}%
\usepackage{amsmath,latexsym}
\usepackage{graphicx}
\usepackage{amsmath}
\usepackage{amsfonts}
\usepackage{amssymb}%
\begin{document}

\title{{Noncommutative-geometry wormholes with
   isotropic pressure}}
   \author{
Peter K. F. Kuhfittig*\\  \footnote{kuhfitti@msoe.edu}
 \small Department of Mathematics, Milwaukee School of
Engineering,\\
\small Milwaukee, Wisconsin 53202-3109, USA}

\date{}
 \maketitle

\begin{abstract}\noindent
The strategy adopted in the original
Morris-Thorne wormhole was to retain
complete control over the geometry at
the expense of certain engineering
considerations.  The purpose of this
paper is to obtain several complete
wormhole solutions by assuming a
noncommutative-geometry background
with a concomitant isotropic-pressure
condition.  This condition allows us to
consider a cosmological setting with a
perfect-fluid equation of state.  An
extended form of the equation
generalizes the first solution and
subsequently leads to the generalized
Chaplygin-gas model and hence to a
third solution.  The solutions
obtained extend several previous
results.  This paper also reiterates
the need for a noncommutative-geometry
background to account for the enormous
radial tension that is characteristic
of Morris-Thorne wormholes.    
\\
\emph{PACS (2020): 04.20-q; 04.20.Jb}

\end{abstract}

\section{Introduction}\label{E:introduction}

Wormholes are tunnel-like structures in
spacetime that connect widely separated
regions of our Universe or different universes
altogether.  Although not entirely new, a
detailed analysis of traversable wormholes
was first performed by  Morris and Thorne
\cite{MT88} in 1988.  They had proposed the
following static and spherically symmetric
line element for a wormhole spacetime:
\begin{equation}\label{E:line1}
ds^{2}=-e^{2\Phi(r)}dt^{2}+\frac{dr^2}{1-b(r)/r}
+r^{2}(d\theta^{2}+\text{sin}^{2}\theta\,
d\phi^{2}),
\end{equation}
using units in which $c=G=1$.  Here $b=b(r)$
is called the \emph{shape function} and
$\Phi=\Phi(r)$ is called the \emph{redshift
function}; $\Phi(r)$ must be everywhere finite
to avoid an event horizon.  The shape function
must also have certain properties, including
the fixed-point property $b(r_0)=r_0$, where
$r=r_0$ is the radius of the \emph{throat} of
the wormhole.  An important requirement is the
\emph{flare-out condition} at the throat:
$b'(r_0)<1$, while $b(r)<r$ near the throat.
For the wormhole spacetime as a whole, the most
important physical property is asymptotic
flatness, which demands that  $\text{lim}
_{r\rightarrow\infty}\Phi(r)=0$ and
$\text{lim}_{r\rightarrow\infty}b(r)/r=0$.

The flare-out condition refers to the
tunnel-like shape of $b(r)$ when viewed, for
example, in an embedding diagram \cite{MT88}.
This condition can only be met by violating
the null energy condition (NEC)
\begin{equation}
  T_{\alpha\beta}k^{\alpha}k^{\beta}\ge 0
\end{equation}
for all null vectors $k^{\alpha}$, where
$T_{\alpha\beta}$ is the energy-momentum
tensor.  Matter that violates the NEC is
called ``exotic" in Ref. \cite{MT88}.  In
particular, for the outgoing null vector
$(1,1,0,0)$, the violation has the form
\begin{equation}
   T_{\alpha\beta}k^{\alpha}k^{\beta}=
   \rho +p_r<0.
\end{equation}
Here $T^t_{\phantom{tt}t}=-\rho$ is the energy
density, $T^r_{\phantom{rr}r}= p_r$ is the
radial pressure, and
$T^\theta_{\phantom{\theta\theta}\theta}=
T^\phi_{\phantom{\phi\phi}\phi}=p_t$ is
the lateral pressure.

Regarding the theoretical construction of a
wormhole, Morris and Thorne adopted the
following stategy: specify the functions
$b=b(r)$ and $\Phi=\Phi(r)$ to produce the
desired geometric properties.  This strategy
retains complete control over the geometry
but leads to enormous practical problems:
the members of the engineering team must
manufacture or search the Universe for
matter or fields that yield the required
energy-momentum tensor.  There are
theoretical problems as well, as can be
seen from the Einstein field equations,
listed next.
\begin{equation}\label{E:Einstein1}
  \rho(r)=\frac{b'}{8\pi r^2},
\end{equation}
\begin{equation}\label{E:Einstein2}
   p_r(r)=\frac{1}{8\pi}\left[-\frac{b}{r^3}+
   2\left(1-\frac{b}{r}\right)\frac{\Phi'}{r}
   \right],
\end{equation}
\begin{equation}\label{E:Einstein3}
   p_t(r)=\frac{1}{8\pi}\left(1-\frac{b}{r}\right)
   \left[\Phi''-\frac{b'r-b}{2r(r-b)}\Phi'
   +(\Phi')^2+\frac{\Phi'}{r}-
   \frac{b'r-b}{2r^2(r-b)}\right].
\end{equation}
Since Eq. (\ref{E:Einstein3}) can be obtained
from the conservation of the stress-energy tensor
$T^{\mu\nu}_{\phantom{\mu\nu};\nu}=0$, only
Eqs. (\ref{E:Einstein1}) and (\ref{E:Einstein2})
are actually needed..  These can be written in
the following forms:
\begin{equation}\label{E:E1}
  b'=8\pi\rho r^2,
\end{equation}
and
\begin{equation}\label{E:E2}
  \Phi'=\frac{8\pi p_rr^3+b}{2r(r-b)}.
\end{equation}
It now becomes apparent that due to the
condition $b(r_0)=r_0$, $\Phi'(r)$ and hence
$\Phi(r)$ are not likely to exist.  (There are
exceptions, however, for certain special
forms of $b(r)$, as shown by Lobo \cite{fL05}.)

The purpose of this paper is to obtain several
complete wormhole solutions by assuming a
noncommutative-geometry background in
conjunction with an isotropic-pressure
condition, discussed in the next section.

\section{Noncommutative geometry}

An important outcome of string theory is
the realization that coordinates may become
noncommutative operators on a $D$-brane
\cite{eW96, SW99}.  Noncommutativity replaces
point-like objects by smeared objects
\cite{SS03, NSS06, NS10} with the aim of
eliminating the divergences that invariably
appear in general relativity.  As a
consequence, spacetime can be encoded in the
commutator $[\textbf{x}^{\mu},\textbf{x}^{\nu}]
=i\theta^{\mu\nu}$, where $\theta^{\mu\nu}$
is an antisymmetric matrix that determines the
fundamental cell discretization of spacetime
in the same way that Planck's constant
discretizes phase space \cite{NSS06}.  The
smearing can be modeled using a Gaussian
distribution of minimal length $\sqrt{\beta}$
instead of the Dirac delta function \cite{NSS06,
mR11, RKRI12, pK13}.  An equally effective way
is to assume that the energy density of the
static and spherically symmetric and
particle-like gravitational source is
given by \cite{NM08, LL12}
\begin{equation}\label{E:rho}
  \rho(r)=\frac{\mu\sqrt{\beta}}
     {\pi^2(r^2+\beta)^2},
\end{equation}
which can be interpreted to mean that
the gravitational source causes the mass
$\mu$ of a particle to be diffused
throughout the region of linear dimension
$\sqrt{\beta}$ due to the uncertainty;
so $\sqrt{\beta}$ has units of length.
Here it is important to note that the
noncommutative effects can be implemented
in the Einstein field equations by
modifying only the energy-momentum tensor,
while leaving the Einstein tensor unchanged.
The reason is that, according to Ref.
\cite{NSS06}, a metric field is a geometric
structure defined over an underlying
manifold.  Its strength is measured by
curvature, which is a response to the
presence of a mass-energy distribution.
Here the key observation is that
noncommutativity is an intrinsic property
of spacetime (rather than a superimposed
geometric structure), thereby affecting
the mass-energy and momentum distributions.
But the energy-momentum density, in turn,
determines the spacetime curvature, which
explains why the Einstein tensor can be
left unchanged.  An important consequence
is that the length scales can be
macroscopic.  We can therefore use Eq.
(\ref{E:rho}) to determine the mass
distribution
\begin{equation}\label{E:mass}
   \int^r_04\pi (r')^2\rho(r')
   dr'= \frac{2M}{\pi}\left(\text{tan}^{-1}
  \frac{r}{\sqrt{\beta}}
  -\frac{r\sqrt{\beta}}{r^2+\beta}\right),
\end{equation}
where $M$ is now the total mass of the source.

Suppose we return the the conservation
law $T^{\alpha}
_{\phantom{\beta r}\gamma;\,\alpha}=0$.  If
$\gamma =r$, then we obtain
\begin{equation}
  \frac{\partial}{\partial r}T^r_{\phantom{tt}r}
  =-\frac{1}{2}g^{tt}\frac{\partial g_{tt}}
  {\partial r}
  (T^r_{\phantom{tt}r}-T^t_{\phantom{tt}t})-
  g^{\theta\theta}\frac{\partial g_{\theta\theta}}
  {\partial r}
  (T^r_{\phantom{tt}r}-
  T^\theta_{\phantom{tt}\theta}).
\end{equation}
According to Ref. \cite{NSS06}, to preserve
the property $g_{tt}=-g^{-1}_{rr}$, we require
that $T^r_{\phantom{tt}r}=T^t_{\phantom{tt}t}=
-\rho(r)$, while
\begin{equation}
   T^\theta_{\phantom{tt}\theta}=
   T^\phi_{\phantom{tt}\phi} =-\rho(r)
   -\frac{r}{2}\frac{\partial\rho(r)}
   {\partial r}.
\end{equation}
Furthermore, a massive structureless point
is replaced by a self-gravitating droplet
of anisotropic fluid of density $\rho$,
which yields the radial pressure
\begin{equation}\label{E:radial}
   p_r(r)=-\rho(r),
\end{equation}
thereby preventing the collapse to a
matter point.  The tangential pressure
is given by
\begin{equation}\label{E:tangential}
    p_t(r)=-\rho(r)-\frac{r}{2}
    \frac{\partial\rho(r)}{\partial r}.
\end{equation}
Since the length scales can be macroscopic,
we can retain Eq. (\ref{E:radial}) and then
use Eq. (\ref{E:tangential}) to determine
\begin{equation}
   p_t(r)=-\rho(r)-\frac{r}{2}
    \frac{\partial\rho(r)}{\partial r}=
    p_r(r)+\frac{2\mu r^2\sqrt{\beta}}
    {\pi^2(r^2+\beta)^3}
\end{equation}
by Eq. (\ref{E:rho}).  So for larger $r$, we
have $p_t(r)\approx p_r(r)$.  Since the
pressure becomes isotropic, we can write
the equation of state in the form
\begin{equation}\label{E:isotropic}
   p(r)=-\rho(r).
\end{equation}
In the present context, this is an important
observation since Morris-Thorne wormholes
are normally characterized by an anisotropic
pressure.

\section{A wormhole solution}\label{S:wormhole}

The conservation law
$T^{\mu\nu}_{\phantom{\mu\nu};\nu}=0$
yields the following equation:
\begin{equation}
   p'_r=-\Phi'\rho-\Phi' p_r-\frac{2}{r}p_r
   +\frac{2}{r}p_t.
\end{equation}
From Eq. (\ref{E:isotropic}) we obtain
\begin{equation}\label{E:main}
   p'=-(p+\rho)\Phi'.
\end{equation}
Given the isotropic pressure, we can safely
assume the perfect-fluid equation of state
\begin{equation}\label{E:perfect}
   p=\omega\rho,
\end{equation}
which, in turn, allows us to consider a
cosmological setting.  Moreover, in Sec.
\ref{S:general} we are going to use the more
general equation of state
\begin{equation}\label{E:generalized}
   \rho=\frac{1/|\omega|}{p^{\alpha}};
\end{equation}
if $\alpha=-1$, we obtain $p=|\omega|\rho$,
which is a special case of Eq. (\ref{E:perfect}).
Eq. (\ref{E:generalized}) not only generalizes
the equation of state, it can be adapted to a
cosmological model usually referred to as
generalized Chaplygin gas, discussed in Sec.
\ref{S:Chaplygin}.

Returning now to Eq. (\ref{E:perfect}), if we
substitute $p=\omega\rho$ in Eq. (\ref{E:main}),
we find that $\omega\rho'=-\rho(1+\omega)\Phi'$
and hence
\begin{equation}\label{E:diffeq}
   \Phi'=-\frac{\omega}{1+\omega}
      \frac{\rho'}{\rho}.
\end{equation}
Integrating, we obtain the redshift function
\begin{equation*}
  \Phi=-\frac{\omega}{1+\omega}(\text{ln}\,
  \rho+\text{ln}\,c)=-\frac{\omega}{1+\omega}
     \text{ln}\,\,c\rho,
\end{equation*}
where $c$ is an arbitrary positive constant.
Substituting $\rho(r)$ from Eq. (\ref{E:rho})
yields
\begin{equation*}
   \Phi=-\frac{\omega}{1+\omega}\,\text{ln}\,
   \frac{c\mu\sqrt{\beta}}{\pi^2(r^2+\beta)^2};
\end{equation*}
so
\begin{equation}\label{E:redshift1}
  e^{2\Phi(r)}=\left[\frac{c\mu\sqrt{\beta}}
  {\pi^2(r^2+\beta)^2}\right]^
  {-\frac{2\omega}{1+\omega}}.
\end{equation}

The next step is to determine the shape
function from Eqs. (\ref{E:E1}) and
(\ref{E:rho}):
\begin{multline}\label{E:shape}
   b(r)=\int^r_{r_0}8\pi(r')^2\rho(r')dr'\\
   =\frac{4m\sqrt{\beta}}{\pi}
  \left(\frac{1}{\sqrt{\beta}}\text{tan}^{-1}
  \frac{r}{\sqrt{\beta}}-\frac{r}{r^2+\beta}-
  \frac{1}{\sqrt{\beta}}\text{tan}^{-1}
  \frac{r_0}{\sqrt{\beta}}+\frac{r_0}{r_0^2
  +\beta}\right)+r_0;
\end{multline}
observe that $b(r_0)=r_0$, as required.  Here
$m$ is the total mass of the source; so $b(r)$ 
is the mass distribution covering both sides 
of the throat.

One of the requirements of a valid wormhole
solution is asymptotic flatness.  For the
shape function, we evidently have
$\text{lim}_{r\rightarrow\infty}b(r)/r=0$.
In Eq. (\ref{E:redshift1}), we are faced with
a more complicated stuation: if $-1<\omega<0$,
then  $\text{lim}_{r\rightarrow\infty}e^{2\Phi}
=0$; if $\omega<-1$ or $\omega\ge 0$, then
$\text{lim}_{r\rightarrow\infty}e^{2\Phi}=
+\infty$.  Either way, our wormhole spacetime
is not asymptotically flat.  Accordingly, the
wormhole material must be cut off at some
$r=a$ and joined to the exterior Schwarzschild
solution
\begin{equation}\label{E:line1}
ds^{2}=-\left(1-\frac{2M}{r}\right)dt^{2}
+\frac{dr^2}{1-2M/r}
+r^{2}(d\theta^{2}+\text{sin}^{2}\theta\,
d\phi^{2}).
\end{equation}
From the shape function $b=b(r)$, we have
\begin{equation}
    \frac{1}{1-\frac{b(a)}{a}}=
        \frac{1}{1-\frac{2M}{a}}.
\end{equation}
So the effective mass of the wormhole is
$M=\frac{1}{2}b(a)$.  For the redshift
function, we then obtain
\begin{equation}
    e^{2\Phi(a)}=1-\frac{2M}{a}=
    1-\frac{b(a)}{a}.
\end{equation}
This junction condition now yields the
constant $c$ by letting $r=a$ in Eq.
(\ref{E:redshift1}):
\begin{equation}
  e^{2\Phi(a)}=\left[\frac{c\mu\sqrt{\beta}}
  {\pi^2(a^2+\beta)^2}\right]^
  {-\frac{2\omega}{1+\omega}}
  =1-\frac{b(a)}{a}.
\end{equation}
Solving for $c$, we have
\begin{equation}
     c=\frac{\pi^2(a^2+\beta)^2}
     {\mu\sqrt{\beta}}\left(1-\frac{b(a)}{a}
     \right)^{-\frac{1+\omega}{2\omega}}.
\end{equation}
Substituting in Eq. (\ref{E:redshift1}) and
simplifying yields the final form
\begin{equation}\label{E:redshift2}
   e^{2\Phi(r)}=\left(\frac{a^2+\beta}
   {r^2+\beta}\right)^{-\frac{4\omega}{1+\omega}}
   \left(1-\frac{b(a)}{a}\right).
\end{equation}
Observe that at $r=a$, we have indeed
\begin{equation}
    e^{2\Phi(a)}=1-\frac{b(a)}{a}=
       1-\frac{2M}{a}.
\end{equation}

The complete solution can now be written as
\begin{equation}\label{E:interior}
   ds^{2}=-e^{2\Phi(r)}dt^{2}+\frac{dr^2}{1-b(r)/r}
 +r^{2}(d\theta^{2}+\text{sin}^{2}\theta\,
 d\phi^{2}),
\end{equation}
for $r\le a$; here $e^{2\Phi(r)}$ is given by
Eq. (\ref{E:redshift2}) and $b(r)$ by Eq.
(\ref{E:shape}).  For $r>a$,
\begin{equation}\label{E:exterior}
   ds^{2}=-\left(1-\frac{b(a)}{r}\right)dt^{2}
   +\frac{dr^2}{1-b(a)/r}
 +r^{2}(d\theta^{2}+\text{sin}^{2}\theta\,
 d\phi^{2}).
\end{equation}

Since the interior solution, Eq.
(\ref{E:interior}), extends only from $r=r_0$
to $r=a$, the solution is valid for all $\omega$,
except for $\omega=-1$, as can be seen from
Eq. (\ref{E:redshift2}).

Returning to the junction surface $r=a$, while
our metric is continuous at $r=a$, the
derivatives may not be.  This behavior needs
to be taken into account when discussing the
surface stresses.  The following forms,
proposed by Lobo \cite{fL05, fL04}, are
suitable for this purpose:
\begin{equation}
   \sigma=-\frac{1}{4\pi a}
   \left(\sqrt{1-\frac{2M}{a}}
   -\sqrt{1-\frac{b(a)}{a}}\right)
\end{equation}
and
\begin{equation}\label{E:junction}
  \mathcal{P}=\frac{1}{8\pi a}
  \left(\frac{1-\frac{M}{a}}
  {\sqrt{1-\frac{2M}{a}}}
  -[1+a\Phi'(a)]\sqrt{1-\frac{b(a)}{a}}
  \right).
\end{equation}
Since $b(a)=2M$, the surface stress-energy
$\sigma$ is zero.  From Eq.
(\ref{E:redshift2}), we obtain
\begin{equation}
   \Phi'(a)=\frac{\omega}{1+\omega}
     \frac{4a}{a^2+\beta}.
\end{equation}
Substituting in Eq. (\ref{E:junction})
and simplifying, we get
\begin{equation}
    \mathcal{P}=\frac{1}{8\pi a\sqrt{1-\frac{2M}{a}}}
    \left[\frac{M}{a}+\frac{\omega}{1+\omega}
    \frac{4a^2}{a^2+\beta}\left(-1+\frac{b(a)}{a}\right)
    \right].
\end{equation}
Due to the condition $b(r)<r$ near the throat,
$\mathcal{P}$ is positive whenever
\begin{equation}
   -1<\omega\le 0.
\end{equation}

\section{The generalized equation of
      state}\label{S:general}

In this section, we return to Eq.
(\ref{E:generalized}), restated here for convenience:
\begin{equation}\label{E:EoS1}
   \rho=\frac{1/|\omega|}{p^{\alpha}},
      \quad \alpha>-1.
\end{equation}
From Eq. (\ref{E:main}),
\begin{equation}
    \Phi'=-\frac{p'}{p+\rho}=
       -\frac{p^{\alpha}p'}{p^{\alpha +1}
       +\frac{1}{|\omega|}}.
\end{equation}
Integrating, we obtain
\begin{equation}
   2\Phi=-\frac{2}{\alpha +1}\,
   \text{ln}\left[c\left(p^{\alpha +1} +
   \frac{1}{|\omega|}\right)\right]
\end{equation}
and hence
\begin{equation}\label{E:redshift3}
   e^{2\Phi}=\left[c\left(p^{\alpha+1}+
   \frac{1}{|\omega|}\right)\right]
   ^{-\frac{2}{\alpha +1}},
\end{equation}
where
\begin{equation}\label{E:pressure}
   p^{\alpha +1}=\left(\frac{\pi^2(r^2+\beta)^2}
   {|\omega|\mu\sqrt{\beta}}\right)
   ^{\frac{\alpha +1}{\alpha}}
\end{equation}
from Eq. (\ref{E:rho}).  As before, the
resulting wormhole spacetime is not
asymptotically flat and needs to be joined
to an external Schwarzschild spacetime, i.e.,
\begin{equation}
   \left[c\left(p(a)^{\alpha +1}
   +\frac{1}{|\omega|}\right)\right]
   ^{-\frac{2}{\alpha +1}
   }=1-\frac{b(a)}{a}.
\end{equation}
Solving for $c$, we get
\begin{equation}
   c=\frac{\left(1-\frac{b(a)}{a}\right)
   ^{-\frac{\alpha +1}{2}}}
   {p(a)^{\alpha +1}+\frac{1}{|\omega|}}
\end{equation}
to be substituted into Eq.
(\ref{E:redshift3}).  The result is
\begin{equation}
   e^{2\Phi(r)}=\left[\frac{p(r)^{\alpha +1}
   +\frac{1}{|\omega|}}
   {p(a)^{\alpha +1}+\frac{1}{|\omega|}}
   \right]^{-\frac{2}{\alpha +1}}
   \left(1-\frac{b(a)}{a}\right).
\end{equation}
Using Eq. (\ref{E:pressure}) and reducing, we
obtain the final form
\begin{equation}\label{E:redshift4}
  e^{2\Phi(r)}=\left[\frac{\left(\frac{\pi^2(r^2+\beta)^2}
  {\mu\sqrt{\beta}}\right)^{\frac{\alpha +1}{\alpha}}
  |\omega|^{-\frac{1}{\alpha}}+1}
  {\left(\frac{\pi^2(a^2+\beta)^2}
  {\mu\sqrt{\beta}}\right)^{\frac{\alpha +1}{\alpha}}
  |\omega|^{-\frac{1}{\alpha}}+1}\right]
  ^{-\frac{2}{\alpha +1}}\left(1-\frac{b(a)}{a}\right).
\end{equation}
So $e^{2\Phi(a)}=1-\frac{b(a)}{a}$.  The shape
function is still given by Eq. (\ref{E:shape}).

We saw earlier that if $\alpha=-1$ in Eq.
(\ref{E:EoS1}), we recover $p=|\omega|\rho$,
a special form of the perfect-fluid equation,
but $e^{2\Phi(r)}$ in Eq. (\ref{E:redshift4})
is not defined for $\alpha=-1$.  However, the
first factor on the right takes on the
indeterminate form $1^{\infty}$ at $\alpha=-1$.
So $\text{lim}_{r\rightarrow -1+}e^{2\Phi(r)}$
can be evaluated by means of L'Hospital's rule.
The result is
\begin{equation}
   e^{2\Phi(r)}=\left(\frac{a^2+\beta}
   {r^2+\beta}\right)
   ^{-\frac{4||\omega|}{1+|\omega|}}
   \left(1-\frac{b(a)}{a}\right),
\end{equation}
which is consistent with our previous form,
Eq. (\ref{E:redshift2}).

\section{The generalized Chaplygin-gas
       model}\label{S:Chaplygin}
Another equation of state of interest to us is
\begin{equation}\label{E:EoS2}
   p=\frac{-K}{\rho^{\alpha}}, \quad
       0<\alpha\le 1.
\end{equation}
The model with this equation of state is
referred to as generalized Chaplygin gas
\cite{fL06, KMP01, BBS02, pK08}.
Cosmologists became interested in this form
of matter when it turned out to be a
candidate for combining dark matter and
dark energy.  To support a wormhole, we
must have
\begin{equation}
   K<\frac{1}{(8\pi r_0^2)^{\alpha +1}}.
\end{equation}

Suppose we rewrite Eq. (\ref{E:EoS1}) in
the form
\begin{equation}
   p=\frac{(1/|\omega|)^{1/\alpha}}
      {\rho^{1/\alpha}}.
\end{equation}
Now assume that $\alpha\ge 1$ and let
$\alpha_1=1/\alpha$.  Then
\begin{equation}
    p=\frac{(1/|\omega|)^{1/\alpha}}
      {\rho^{\alpha_1}}, \quad
         0<\alpha_1\le 1.
\end{equation}
If we now choose
\begin{equation}
   K=-\left(\frac{1}{|\omega|}\right)
   ^{1/\alpha}=-|\omega|^{-\frac{1}{\alpha}},
\end{equation}
we have a valid equation of state for
the Chaplygin model since $K$ can be
determined from the free parameter
$\omega$.

It now follows directly from Eq.
(\ref{E:redshift4}) that for the generalized
Chaplygin model,
\begin{equation}
  e^{2\Phi(r)}=\left[\frac{\left(\frac{\pi^2(r^2+\beta)^2}
  {\mu\sqrt{\beta}}\right)^{\alpha_1 +1}K-1}
  {\left(\frac{\pi^2(a^2+\beta)^2}
  {\mu\sqrt{\beta}}\right)^{\alpha_1 +1}K-1}\right]
  ^{-\frac{2\alpha_1}{\alpha_1 +1}}\left(1-\frac{b(a)}{a}\right).
\end{equation}
As before, $e^{2\Phi(a)}=1-\frac{b(a)}{a}.$
The shape function is again given by
Eq. (\ref{E:shape}), thereby producing a
complete wormhole solution.

\section{More on noncommutative geometry: the
   radial \\tension}\label{S:more}

While noncommutative geometry was
instrumental in obtaining a wormhole solution,
its role may extend well beyond the present
study.  For example, there is an aspect of
wormhole physics that goes back to Ref.
\cite{MT88} but has not been discussed so
far: recalling that the radial tension
$\tau(r)$ is the negative of $p_r(r)$
and reintroducing $c$ and $G$, the radial
tension is given by \cite{MT88}
\begin{equation}\label{E:tau}
   \tau(r)=\frac{b(r)/r-2[r-b(r)]\Phi'(r)}
   {8\pi Gc^{-4}r^2}.
\end{equation}
From this condition it follows that the
radial tension at the throat is
\begin{equation}\label{E:tension1}
   \tau=\frac{1}{8\pi Gc^{-4}r_0^2}\approx
   5\times 10^{41}\frac{\text{dyn}}{\text{cm}^2}
   \left(\frac{10\,\text{m}}{r_0}\right)^2.
\end{equation}
In particular, for $r_0=3\,\,\text{km}$,
$\tau$ has the same magnitude as the pressure
at the center of a massive neutron star.
Attributing this outcome to exotic matter
is problematical at best since exotic matter
was introduced for a completely different
reason: matter is called exotic if it
violates the NEC.

It was shown in a recent paper (Kuhfittig
\cite{pK20}) that noncommutative geometry
can account for the large radial tension.
As an offshoot of string theory, it can
therefore be viewed as a foray into
quantum gravity, thereby going beyond
classical general relativity.

\section{Conclusions}

The strategy for the theoretical construction
of wormholes adopted by Morris and Thorne
\cite{MT88} was to specify the desired
geometric properties of the wormhole and
then manufacture or search the Universe for
matter or fields that would produce the
corresponding energy-momentum tensor.  The
purpose of this paper is to determine
complete wormhole solutions by starting with
a noncommutative-geometry background.  Given
that the noncommutative effects can be
implemented in the Einstein field equations
without changing the Einstein tensor, it
follows that the length scales can be
macroscopic.  It was also concluded that
for larger $r$, the radial and transverse
pressures become equal, thereby yielding
the equation of state $p=\omega\rho$,
allowing us to consider a cosmological
setting.  This case is discussed in Sec.
\ref{S:wormhole}.  The solution obtained
is not asymptotically flat and needs to
be cut off and joined to an exterior
Schwarzschild solution at the junction
surface $r=a$.  It was found that the
solution is valid for all $\omega\neq
-1$.  (An examination of the surface
stresses has shown, however, that a
positive surface pressure at the
junction surface is obtained only if
$-1<\omega\le 0$.)

The last part of the paper uses the
more general equation of state $\rho
=(1/|\omega|)p^{-\alpha}$, $\alpha>-1$.
This equation of state not only
generalizes the previous solution,
it can also be adapted to another
cosmological model, generalized
Chaplygin gas.  The result is two
additional complete wormhole solutions.

Previous studies \cite{fL05, sS05}
have shown that wormhole solutions
exist for $\omega<-1$ since, in a
cosmological setting, we are dealing
with phantom energy, which is known
to violate the null energy condition.
While fulfilling the primary
prerequisite for the existence of
wormholes, this approach yields
neither the shape nor the redshift
functions.  Similarly, a
noncommutative-geometry background
alone is sufficient for determining
$b=b(r)$, as shown in Ref.
\cite{pK15}, but it does not yield
the redshift function $\Phi=
\Phi(r)$.

As a final remark, Sec. \ref{S:more}
reiterates another aspect of
noncommutative geometry, the ability
to account for the enormous radial
tension in a typical traversable
wormhole, discussed in Ref.
\cite{pK20}.

\end{document}